\documentclass[atoms,article,accept,moreauthors,pdftex,10pt,a4paper]{Definitions/mdpi}
\pdfoutput=1
\firstpage{1} 
\pubvolume{1}
\issuenum{1}
\articlenumber{0}
\pubyear{2022}
\copyrightyear{2022}
\externaleditor{Academic Editor: Emmanouil P. Benis} 
\datereceived{} 
\dateaccepted{} 
\datepublished{} 
\hreflink{https://doi.org/} % If needed use \linebreak
\def\physscr{Phys.~Scr.}

\newcommand{\abun}{$\log(N_{\rm el}/N_{\rm tot})$}

\def\ione{\,{\sc i}}
\def\ii{\,{\sc ii}}

\def\vmicro{$\xi_{\rm t}$}

\newcommand{\kms}{km\,s$^{-1}$}
\newcommand{\te}{$T_{\rm eff}$}
\newcommand{\logg}{$\log{g}$}
\newcommand{\vsini}{$v\sin{i}$}
\def\vmacro{$\zeta_{\rm RT}$}

\newcommand{\SME}{\textsc{sme}}

\Title{Using Molecular Lines to Determine Carbon and Nitrogen Abundances in the Atmospheres of Cool Stars}

\TitleCitation{Using Molecular Lines to Determine Carbon and Nitrogen Abundances in the Atmospheres of Cool Stars}

\Author{Tatiana Ryabchikova %MDPI: Please carefully check the accuracy of names and affiliations.
 $^{1,}$*\orcidA{}, Nikolai Piskunov $^{2}$\orcidB{} and Yury Pakhomov $^{1}$\orcidC{}}

\AuthorNames{Tatiana Ryabchikova, Nikolai Piskunov and Yury Pakhomov}

\AuthorCitation{Ryabchikova, T.; Piskunov, N.; Pakhomov, Y.}

\address{%
$^{1}$ \quad Institute of Astronomy, Russian Academy of Sciences, 119017 Moscow, Russia;  pakhomov@inasan.ru \\ %please add post code
$^{2}$ \quad Department of Physics and Astronomy, Division of Astronomy and Space Physics, Uppsala University, P.O. Box 516, 751 20 Uppsala, Sweden; nikolai.piskunov@physics.uu.se} %please add post code

\corres{Correspondence: ryabchik@inasan.ru; Tel.: +7-495-951-3980}

\abstract{Simultaneous analysis of the C$_2$ and CN molecular bands in the 5100--5200 and 7930--8100~\AA\ spectral regions is a promising alternative for the accurate determination of the carbon (C) and nitrogen (N) abundance in the atmospheres of the solar-like stars. Practical implementation of this new method became possible after recent improvements of the molecular constants for both molecules. The new molecular data predicted the correct line strength and line positions; therefore, they were included in the Vienna Atomic Line Database (\textsc{VALD}), which is widely used by astronomers and spectroscopists. In this paper, we demonstrate that the molecular data analysis provides C and, in particular, N abundances consistent with those derived from the atomic lines. We illustrate this by performing the analysis for three stars. Our results provide strong arguments for using the combination of C$_2$ and CN molecular lines for accurate nitrogen abundance determination keeping in mind the difficulties of using the N\ione\ lines in the observed spectra of the solar-like stars.}


\begin{document}
%%%%%%%%%%%%%%%%%%%%%%%%%%%%%%%%%%%%%%%%%%
%% Only for the journal Gels: Please place the Experimental Section after the Conclusions

%%%%%%%%%%%%%%%%%%%%%%%%%%%%%%%%%%%%%%%%%%
%\setcounter{section}{-1} %% Remove this when starting to work on the template.
%\section{How to Use this Template}
%The template details the sections that can be used in a manuscript. Note that the order and names of article sections may differ from the requirements of the journal (e.g., the positioning of the Materials and %Methods section). Please check the instructions for authors page of the journal to verify the correct order and names. For any questions, please contact the editorial office of the journal or support@mdpi.com. %For LaTeX related questions please contact Janine Daum at latex-support@mdpi.com.
%The order of the section titles is: Introduction, Materials and Methods, Results, Discussion, Conclusions for these journals: aerospace,algorithms,antibodies,antioxidants,atmosphere,axioms,biomedicines,carbon,crystals,designs,diagnostics,environments,fermentation,fluids,forests,fractalfract,informatics,information,inventions,jfmk,jrfm,lubricants,neonatalscreening,neuroglia,particles,pharmaceutics,polymers,processes,technologies,viruses,vision

\section{Introduction}

{The abundances %MDPI: Please confirm if the bold in main text should be removed. please check all bold in main text
of CNO elements in stellar spectra are extremely important. They reflect many aspects of star formation
and stellar evolution. Recent systematic observational and theoretical studies of planet formation have introduced new
important aspects of stellar enrichment/depletion of these elements through their inclusion in dust particles
(e.g., \citep{2021NatAs...5.1163S,2021A&A...655A..51K}). Differential depletion of certain elements in the accretion
flow due to pebble migration in a circumstellar disk and its evolution throughout the lifetime of the disk may
and probably does leave a subtle imprint on the relative abundances of the volatile versus refractory elements in
stellar atmospheres. The differences between atmospheric abundances of the binary components are very small; therefore, advanced methods for measuring abundances with a
precision of 0.03~dex or better are needed. Such methods have been proposed and implemented for carbon and oxygen.
They use atomic lines, usually affected by deviations from local thermodynamic equilibrium (NLTE effects) in
comparison with abundances derived from diatomic molecules (not subject to strong NLTE effects), such as CH and OH.
Such techniques have become possible due to recent advances in atomic and molecular data that proved to be crucial
for modeling the NLTE effects in atomic lines and improved the accuracy of the position and strength of individual
transitions in molecular bands. For carbon and oxygen, the abundances derived from
molecular lines are in excellent agreement with non-LTE analyses of atomic lines. For nitrogen, however, even
for the sun, the molecular transitions give a 0.12~dex larger abundance than the atomic lines
(\citep{2019A&A...630A.104A,2021A&A...656A.113A},). This is related to the fact that, for solar-like stars, atomic
nitrogen has no convenient lines for the analysis comparable to  the O\ione\ 6156-58, 7771-77, 8446 and
9260-66~\AA\ lines. For example, in the analysis of the nitrogen abundance, 
\citet{2020A&A...636A.120A} were only able to use four atomic lines in the 7400--8700~\AA\, and one line in the
near infrared region. Moreover, in the spectra of solar-like stars, practically all these lines are blended either
with molecular CN lines or with atomic lines of other elements (see Figure~\ref{N1-lines}), which requires
accurate parameters for all contributing lines. Note, that for stars cooler than the sun, it is much more difficult to use N\ione\ atomic lines because they become weaker, while the CN blends become stronger. Even the molecular bands of NH are hard to reach as they are mostly
located in the UV between 2300 and 3400~\AA\ or in the infrared beyond 1.1~$\mu$, while the spectra obtained
for systematic abundance determinations typically cover the range of 3700--9000~\AA\ or at least some parts of it.

\begin{figure}[H]
%	\centering
	\includegraphics[width=0.99\textwidth]{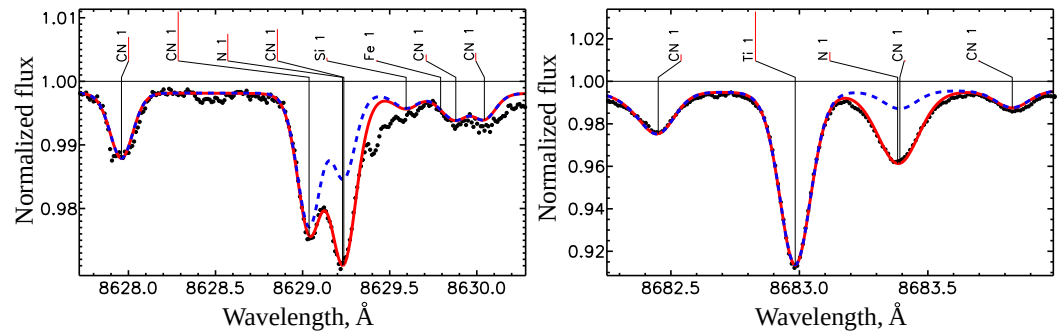}
	\caption{Comparison between the Solar Flux Atlas (black dots) and spectrum synthesis in the regions
	         of N\ione\ atomic lines (red line). The blue dashed line shows the theoretical calculations without the
	         N\ione\ lines demonstrating a blended contribution.} %moved figures after first mentioned, please confirm
	\label{N1-lines}
\end{figure}

The excellent precision of carbon abundance determination for the sun
\citep{2015MNRAS.453.1619A, 2019A&A...624A.111A} motivated us to try to combine the analysis for C and N by
simultaneously interpreting the CN molecular lines around 7950~\AA\ and C$_2$ lines in 5100--5200~\AA\ (Swan band).
In this paper, we present the first results of this analysis. Eventually, the method can be further refined
by including the atomic lines of carbon, NLTE, and 3D effects, but we would like to demonstrate its feasibility
and also illustrate practical aspects, such as the accessibility of the relevant observations, the availability
and the quality of molecular data, and the consistency of the results with other recent determinations.

In the next section we describe our target selection, observational material, and global stellar parameters
determination. In Section \ref{C_N_abun} we present the method and our results, which are then discussed in
Section \ref{Discussion}.

\section{Observations and Stellar Atmospheric Parameters}
\label{Observations}

For the analysis, we selected the sun and the wide binary system 16~Cyg with solar-like components. The spectrum
of the solar light reflected from asteroid Vesta and spectra of the 16~Cyg components were obtained with the
Echelle Spectro Polarimetric Device for Observation of Stars (ESPaDOnS) spectrograph mounted at the 3.6 m
Canada--France--Hawaii Telescope (\citep{2003ASPC..307...41D}, CHFT). The spectral resolving power was R = 81\,000,
and the wavelength coverage was 3696--10482\,\AA. All spectra were downloaded from the Canadian Astronomy Data
Centre (CADC) archive \citep{1994ASPC...61..123C}. The details of the reduction procedure can be found in~\citep{2022MNRAS.514.4958R}. For the sun, we  complemented the Vesta spectrum with the Solar Flux Atlas~\citep{1984sfat.book.....K}, which has much higher spectral resolution and better signal-to-noise ratio
(S/N) than the ESPaDOnS observations.

The fundamental stellar parameters (even for the sun), such as effective temperature \te, surface gravity
\logg, metallicity [M/H], projected equatorial rotation (\vsini), micro- (\vmicro) and macro-turbulent
(\vmacro) velocities as well as the radial velocities were obtained with the help of Spectroscopy Made
Easy (\SME) software package designed for automatic spectral analysis
\citep{1996AAS..118..595V,2017A&A...597A..16P}.
%\SME\ was successfully applied to determinations of the atmospheric parameters and abundances of different stars \citep{2005ApJS..159..141V, 2016ApJS..225...32B}. The accuracy of the atmospheric parameter determinations with \SME\ was investigated by \citet{2016MNRAS.456.1221R} who showed that effective temperatures, surface gravities, and metallicities agree within $\pm65$~K, $\pm0.12$~dex, and $\pm0.04$~dex with other spectroscopic determinations.
\SME\ analysis consists of fitting the synthetic spectrum to the observations in chosen spectral regions.
The stellar atmospheric parameters are the free parameters of the fit, and a set of weights defines the relative
importance of the residuals for each free parameter. When using \SME\, one can select between two approaches:
either work with a handful trusted spectral lines or use all the lines that can be reproduced. The first
approach was more common in the early days of stellar spectroscopy due to the patchy quality and poor
completeness of atomic and molecular data. The second approach assumes that the distribution of errors in the
line data is at least centred on zero and that using many lines will result in correct atmospheric parameters.
This assumption is closer to reality for some parameters (e.g., \te) and further away for the others
(e.g., \logg), but as more and better atomic and molecular data become available through various sources
(data producers, compilations, and data centres), the second approach is gaining momentum.

In our analysis, all atomic and molecular line parameters were extracted from \linebreak{}VALD~\citep{vald2015},
which is one of the nodes of the Virtual Atomic and Molecular Data Centre \textsc{VAMDC}
\citep{2016JPhB...49g4003D,2020Atoms...8...76A}. The line parameters of N\ione\ \textsc{VALD} were collected
from the NIST Atomic Spectra Database \citep{NIST10}. Those for the molecular C$_2$ and CN lines came to VALD
from \citet{BBSB} and \citet{BRW}, respectively. The line parameters are extracted by the users via
VALD extraction tools. In the atmospheres of our target stars, including the sun (Vesta), the atmospheric
parameters were determined by \citet{2022MNRAS.514.4958R} using \textsc{LLmodels} atmospheric grid~\citep{2004AA...428..993S}. For the solar atlas, we did not perform a fit. Instead, we used the
canonical solar 1D atmospheric model computed with the \textsc{MARCS} (Model Atmospheres of Radiative
and Convective Stars) code \citep{2008AA...486..951G}. the atmospheric parameters are presented in
Table~\ref{tab1}.

\begin{table}[H] 
\caption{Stellar parameters derived for the sun (Vesta) and in the binary system 16~Cyg using atomic and
molecular lines.\label{tab1}}
\newcolumntype{C}{>{\centering\arraybackslash}X} %\raggedright 左对齐 \raggedleft 右对齐
\begin{tabularx}{\textwidth}{m{2.6cm}<{\raggedright}CCCC}

%\begin{tabular}{lcccc}
\toprule
\textbf{Parameter}&\textbf{Sun (atlas)}&\textbf{Sun (Vesta)}&\textbf{16~Cyg A}&\textbf{16~Cyg B}\\
\midrule
\te, K        & 5777 & 5778 & 5829 & 5760 \\
\logg, dex    & 4.44 & 4.44 & 4.33 & 4.39 \\
{[M/H]}      & 0.0  & 0.003& 0.110& 0.074\\ 
\vmicro, \kms & 0.90 & 0.86 & 0.99 & 0.90 \\
\vmacro, \kms & 3.50 & 3.59 & 4.21 & 3.32 \\
\bottomrule
\end{tabularx}
\end{table}
\unskip
 
\section{Carbon and Nitrogen Abundance Determination}
\label{C_N_abun}

\citet{2022MNRAS.514.4958R} derived the C and N abundances using atomic (C,N) and molecular (C$_2$) lines in
the atmospheres of the sun (using the Vesta spectrum) and 16~Cyg components. The carbon abundances from
the atomic lines were derived from seven lines accounting for departures from the local thermodynamic
equilibrium as described by \mbox{\citet{Tsymbal2018}}.
Those from the molecules were derived by using the C$_2$ molecular lines
at 5100-5200~\AA\, (the Swan band), using the conventional \SME\ procedure. The nitrogen abundances were 
also estimated using two N\ione\,$\lambda\lambda$~8629, 8683\,\AA\, lines located in the wing of
a strong Ca\ii\, line, a member of the IR-triplet, and partially blended with weak molecular CN lines.
Carbon and nitrogen abundances are collected in Table~\ref{tab2}. Elemental abundances are given as
relative values \abun, which corresponds to $\log\epsilon_{el}$=\abun\,+ 12.04 in the atmospheres with
the solar He abundance. For all stars except the solar atlas, the abundances derived using the atomic lines were
taken from \citet{2022MNRAS.514.4958R}. Carbon and nitrogen solar abundances for the Solar Flux Atlas
analysis were determined by the same set of spectral lines employing the same NLTE abundance determination
procedure \citep{Tsymbal2018} with the NLTE treatment from \citet{2015MNRAS.453.1619A}. The quality of
the NLTE spectrum fitting is illustrated in Figure~\ref{N1-lines}. Note that the carbon abundance derived by using atomic lines based on a 1D static plane-parallel solar model atmosphere, namely
$\log\epsilon_{C}$ = 8.44 $\pm$ 0.04, exactly matched  the state-of-the-art 3D NLTE carbon abundance
determinations \citep{2019A&A...624A.111A}. 
%Note also, that carbon abundances from atomic and molecular lines perfectly agree between each other.       

Although both N\ione\ lines used in our analysis provide consistent abundance results, the abovementioned
difficulties in using them for the abundance analysis of a large set of stars require an independent possibility
for nitrogen abundance determination. Therefore, we performed simultaneous \SME\ analysis of the C$_2$ and CN
molecular lines in two chosen spectral regions, 5100-5200~\AA\ and 7930-8100~\AA. The atmospheric
parameters presented in Table~\ref{tab1} were fixed, while the C and N abundances were allowed to vary. Spectral
masks in both spectral regions were constructed from numerous unblended and slightly blended lines of C$_2$
and CN molecules. We used an \SME\ mask to select 21 intervals in the 5100--5200~\AA\ region and 12 intervals in 7930--8100~\AA\ 
region for the fitting procedure. An example of such a mask for the solar atlas in the 5100--5200~\AA\ region is shown in Figure~\ref{SME}.
The abundance results from the molecular lines are presented in Table~\ref{tab2}, while
Figure~\ref{comp} zooms in on small fragments of the selected spectral intervals to allow better comparison between
the observed and synthesized spectra of our stars.
It is impossible to mark the positions of all individual C$_2$ lines near the head of the Swan band;
therefore, we simply show the synthetic spectra calculated with and without C$_2$ lines.

\begin{figure}[H]
%	\centering
	\includegraphics[width=\textwidth]{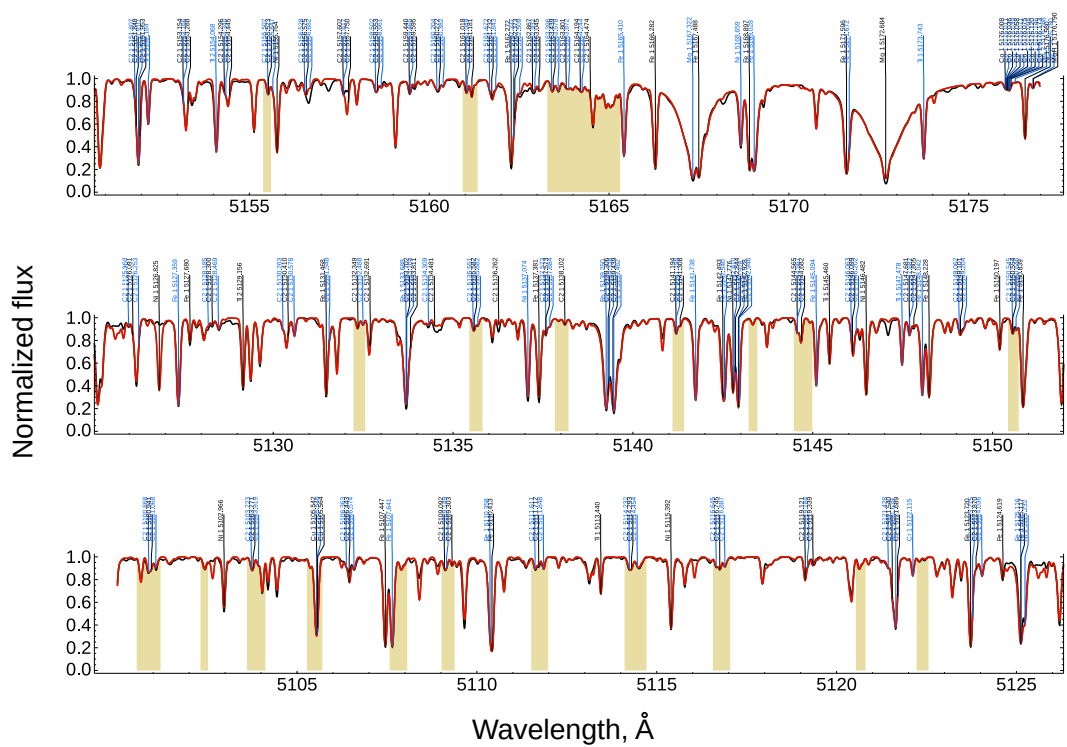}
	\caption{Spectral mask in the 5100--5200~\AA\ region of the solar atlas. Observations are shown by the black line.
	The color bands highlight selected intervals. The \SME\ fitting results are shown by the red line. Only 10\%\ of the strongest
	transitions are marked on the plot.}
	\label{SME}
\end{figure}

The abundance %MDPI: We removed the bold of this paragraph. Please confirm this revision.
uncertainties given in Table~\ref{tab2} are based on the cumulative distribution analysis performed by \SME\ and tend to be
overestimated. In the case of the nitrogen abundance, the statistics (number of points significantly affected by nitrogen abundance) was relatively
small in comparison to the carbon (only CN lines versus CN + C$_2$ lines), and so the core of the distribution is not sufficiently
well-defined leading to a less certain estimate. An alternative estimate based on the diagonal of the covariance matrix, which assumes
the perfect spectral model generated by \SME, gives a similar uncertainty for C abundance and smaller values for N abundance, typically
around 0.04~dex for all stars. 

%A comparison between the observed and synthesized spectra around N\ione\,$\lambda\lambda$8629, 8683\,\AA\, lines are displayed in Figure~\ref{N1-lines}.
    
\begin{figure}[H]
%	\centering
	\includegraphics[width=\textwidth]{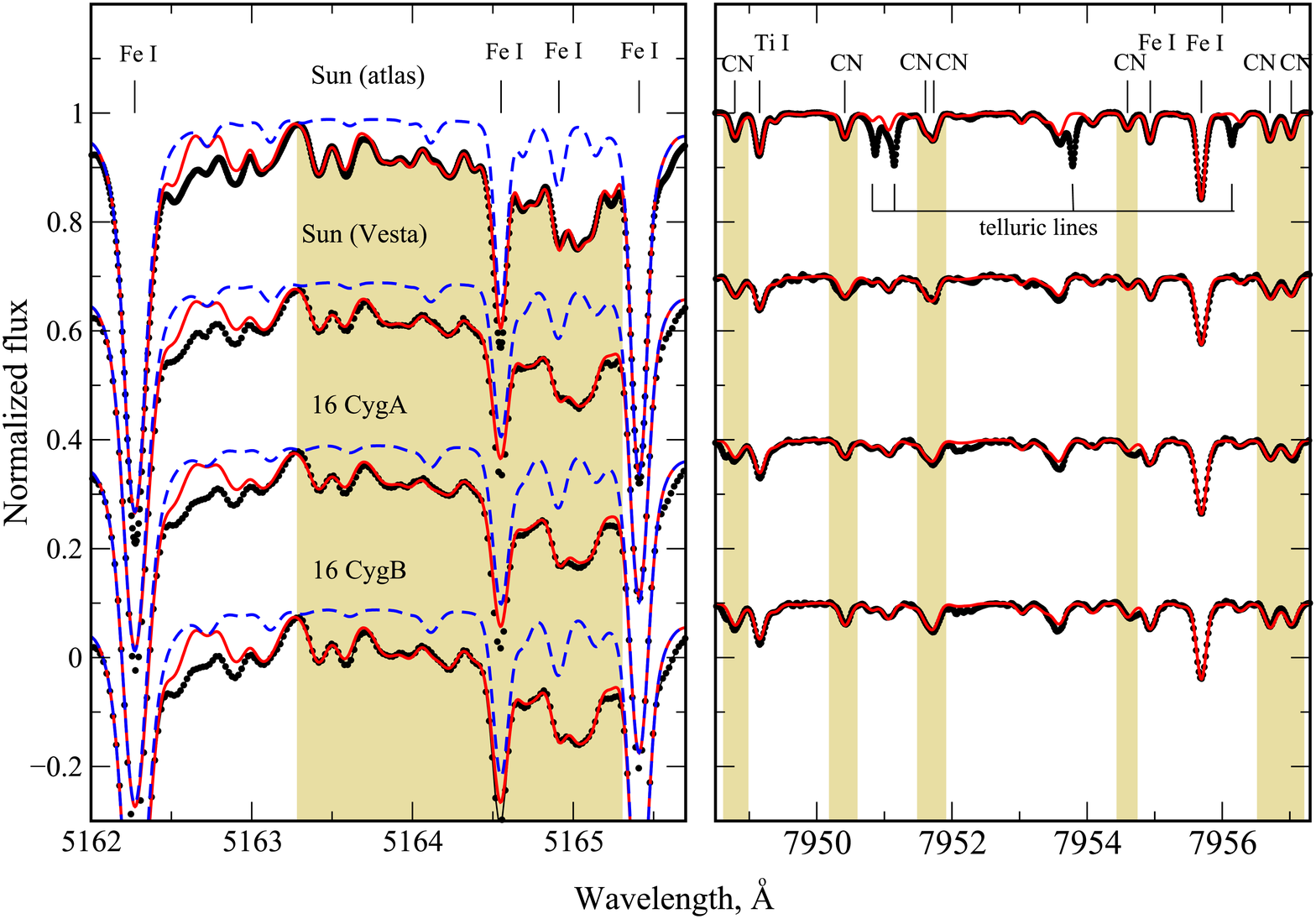}
	\caption{Comparison between the observations and spectrum synthesis in the regions of the
	         C$_2$ Swan bandhead (left panel) and CN molecular lines (right panel). The observations are shown
	         by black dots, the synthetic spectra calculated with the determined atmospheric parameters
	         and abundances are shown by the solid red line. The fitted intervals are marked by colored bands.
	         The blue dashed lines in the left panel demonstrate theoretical calculations without the C$_2$ molecular lines.}
	\label{comp}
\end{figure}\vspace{-6pt}

\begin{table}[H] 
\caption{Carbon and nitrogen abundances in the sun and in the binary system 16~Cyg derived from the
analysis of the atomic and molecular lines.
\label{tab2}}
%\newcolumntype{C}{>{\centering\arraybackslash}X}
%%\newcolumntype{C}
\newcolumntype{C}{>{\centering\arraybackslash}X} %\raggedright 左对齐 \raggedleft 右对齐
\begin{tabularx}{\textwidth}{m{1.6cm}<{\raggedright}CCCC}

%\begin{tabular}{lcccc}
\toprule
\textbf{Species}&\textbf{Sun (atlas)}&\textbf{Sun (Vesta)}&\textbf{16~Cyg A}&\textbf{16~Cyg B}\\
\midrule
C (atom)    &$-$3.596~$\pm$~0.035 &$-$3.601~$\pm$~0.027 &$-$3.560~$\pm$~0.037 &$-$3.564~$\pm$~0.037 \\
C (mol)     &$-$3.600~$\pm$~0.010 &$-$3.617~$\pm$~0.024 &$-$3.564~$\pm$~0.016 &$-$3.571~$\pm$~0.013 \\  %0.010  & 0.003 & 0.008 & 0.007
%C (mol)    &$-$3.600~$\pm$~0.010 &$-$3.622~$\pm$~0.049 &$-$3.570~$\pm$~0.029 &$-$3.554~$\pm$~0.032 \\ 
N (atom)    &$-$4.089~$\pm$~0.010 &$-$4.062~$\pm$~0.033 &$-$4.063~$\pm$~0.009 &$-$4.076~$\pm$~0.017 \\
N (mol)     &$-$4.072~$\pm$~0.043 &$-$4.059~$\pm$~0.084 &$-$4.006~$\pm$~0.070 &$-$4.038~$\pm$~0.058 \\  %0.086  & 0.025 & 0.048 & 0.040
%N (atom_sme)&-4.149$\pm$0.051 &-4.121$\pm$0.089 &-4.024$\pm$0.083 &-4.056$\pm$0.146 \\
\bottomrule
\end{tabularx}
\end{table}
\unskip
%Numbered lists can be added as follows:
%\begin{enumerate}[leftmargin=*,labelsep=4.9mm]
%\item	First item
%\item	Second item
%\item	Third item
%\end{enumerate}

%We also performed \SME\ analysis of the small regions around four N\ione\,$lambda\lambda$~7442.3, 8216.3, 8629.2, 8683.4~\AA\, lines in all stars of our progrramme including blending molecular lines of CN. Carbon abundances were fixed. Nitrogen abundances derived by this way are given in the last line of Table~\ref{tab2}. All lines are weak and blended, with the central intensities less than 2-3~\%, therefore the errors of abundance determinations are relatively large.  
   
%%%%%%%%%%%%%%%%%%%%%%%%%%%%%%%%%%%%%%%%%%
\section{Discussion}
\label{Discussion}

We compared the results derived from the 1D LTE (molecules) and NLTE (atoms) analysis of the
Solar Flux Atlas with the advocated solar abundances recently published 
by \citet{2021A&A...653A.141A}. The C and N solar abundances were mostly based on the methods and
results presented in Amarsi {et al.} \citep{2019A&A...624A.111A, 2020A&A...636A.120A, 2021A&A...656A.113A}.
%MDPI: We removed the bold of the two paragraphs. Please confirm this revision.

The carbon and nitrogen abundances derived from the atomic and molecular lines agreed well
with each other. Our 1D NLTE average carbon abundance 
$\log(N_{\rm C}/N_{\rm tot})$ = $-3.60$~$\pm$~$0.04$ or $\log\epsilon_{N}$~=~8.44~$\pm$~0.04 also agreed with
the advocated solar C abundance 8.46~$\pm$~0.04 based on the 3D NLTE analysis of atomic and molecular
lines in the solar disk-center intensity spectrum.  
\textls[-15]{The resulting nitrogen abundance from the atomic lines was}
$\log(N_{\rm N}/N_{\rm tot})$~=~$-4.09$~$\pm$~$0.01$ or $\log\epsilon_{N}$~=~7.95~$\pm$~0.01. It agreed with
the molecular nitrogen abundance derived by \citet{2021A&A...656A.113A} using the same MARCS solar
model, and it was only 0.06~dex higher than their N abundance from the 3D modelling of the molecular lines. 
The nitrogen abundance derived by using the CN lines was $\log\epsilon_{N}$~=~7.97~$\pm$~0.04. Applying the 3D
correction of $\approx$-0.06~dex from \citet{2021A&A...656A.113A}, we obtained
$\log\epsilon_{N}$~=~7.91~$\pm$~0.04 for the solar N abundance from molecular lines. The 3D correction for the
nitrogen atomic lines was smaller, $\approx$-0.04~dex \citep{2020A&A...636A.120A}, and its application
resulted in a nitrogen abundance of $\log\epsilon_{N}$~=~7.91~$\pm$~0.01. Our results
obtained from the analysis of the Solar Flux spectrum well agreed with the $\log\epsilon_{N}$~=~7.89~$\pm$~0.04
derived by \citet{2021A&A...656A.113A} from the 3D analysis of molecular NH+CN lines in the solar
disk-centre intensity spectrum, and it exceeded by 0.14~dex the nitrogen abundance derived by the same
group \citep{2020A&A...636A.120A} from the atomic lines. 

\textls[-15]{Note that our 'atomic line' N abundance agreed within the error bars with the} $\log\epsilon_{N}$~=~7.88 derived by \citet{2009A&A...498..877C}
in their 3D analysis of the equivalent widths of the N\ione\ lines in the solar disk-center intensity spectrum.
We have to emphasize that the 3D studies, \mbox{\citet{2009A&A...498..877C}} and \citet{2020A&A...636A.120A}, of the N\ione\ atomic
lines were based on the equivalent \begin{center}
	
\end{center}widths of partially blended lines, while our analysis was based on
fitting of the observed line profiles with the blend contribution accounted for when computing the line absorption
coefficient.

It is also interesting that while the equivalent widths in
\citet{2009A&A...498..877C, 2020A&A...636A.120A} were measured using the same observed solar disk-center
intensity spectra, they systematically differed from the smaller values reported in
\citet{2020A&A...636A.120A}.

%Authors should discuss the results and how they can be interpreted in perspective of previous studies and of the working hypotheses. The findings and their implications should be discussed in the broadest context possible. Future research directions may also be highlighted.

%%%%%%%%%%%%%%%%%%%%%%%%%%%%%%%%%%%%%%%%%%
%\section{Materials and Methods}

%Materials and Methods should be described with sufficient details to allow others to replicate and build on published results. Please note that publication of your manuscript implicates that you must make all materials, data, computer code, and protocols associated with the publication available to readers. Please disclose at the submission stage any restrictions on the availability of materials or information. New methods and protocols should be described in detail while well-established methods can be briefly described and appropriately cited.

%Research manuscripts reporting large datasets that are deposited in a publicly available database should specify where the data have been deposited and provide the relevant accession numbers. If the accession numbers have not yet been obtained at the time of submission, please state that they will be provided during review. They must be provided prior to publication.

%Interventionary studies involving animals or humans, and other studies require ethical approval must list the authority that provided approval and the corresponding ethical approval code.

%%%%%%%%%%%%%%%%%%%%%%%%%%%%%%%%%%%%%%%%%%
\section{Conclusions}

Our simultaneous analysis of the spectral regions with the C$_2$ and CN molecular lines in solar-like stars,
based on the accurate calculations of the molecular line parameters collected in \textsc{VALD}, provided C
and N abundances consistent with those derived from the analysis of atomic lines. However, often the
atomic N\ione\ lines are difficult or even impossible to use in nitrogen abundance determinations;
therefore, this task can be solved by the simultaneous analysis of C$_2$ and CN molecular lines.      

%%%%%%%%%%%%%%%%%%%%%%%%%%%%%%%%%%%%%%%%%%
%\section{Patents}
%This section is not mandatory, but may be added if there are patents resulting from the work reported in this manuscript.

%%%%%%%%%%%%%%%%%%%%%%%%%%%%%%%%%%%%%%%%%%
\vspace{6pt}

%%%%%%%%%%%%%%%%%%%%%%%%%%%%%%%%%%%%%%%%%%
%% optional
%\supplementary{The following are available online at www.mdpi.com/link, Figure S1: title, Table S1: title, Video S1: title.}

% \supplementary{The following are available at www.mdpi.com/link: Figure S1: title, Table S1: title, Video S1: title. A supporting video article is available at doi: link.}

%%%%%%%%%%%%%%%%%%%%%%%%%%%%%%%%%%%%%%%%%%
\authorcontributions{T.R. and N.P. performed atomic and molecular line analysis and wrote the main text
of the paper; Y.P. prepared the lists of molecular lines for VALD and created the plots. All authors have read and agreed to the published version of the manuscript.}

\funding{The current research was partially funded by the Ministry of Science and Higher Education of
	the Russian Federation under the grant 075-15-2020-780 (N13.1902.21.0039) for T.R. and Y.P.
	%Please add: ``This research received no external funding'' or ``This research was funded by NAME OF FUNDER grant number XXX.'' and  and ``The APC was funded by XXX''. Check carefully that the details given are accurate and use the standard spelling of funding agency names at \url{https://search.crossref.org/funding}, any errors may affect your future funding.
}

\dataavailability{The A\&M data used in this study are openly available in the VALD database \url{http://vald.inasan.ru/~vald3/php/vald.php}, \url{http://vald.astro.uu.se/~vald/php/vald.php}. The observational and synthetic spectra are available on request from the corresponding author.
	 %In this section, please provide details regarding where data supporting reported results can be found, including links to publicly archived datasets analyzed or generated during the study. Please refer to suggested Data Availability Statements in section ``MDPI Research Data Policies'' at \url{https://www.mdpi.com/ethics}. If the study did not report any data, you might add ``Not applicable'' here.
} 

%%%%%%%%%%%%%%%%%%%%%%%%%%%%%%%%%%%%%%%%%%
\acknowledgments{The facilities
of the Canadian Astronomy Data Centre operated by the National Research Council of Canada with the support
of the Canadian Space Agency were used in the work.}

%%%%%%%%%%%%%%%%%%%%%%%%%%%%%%%%%%%%%%%%%%
\conflictsofinterest{The authors declare no conflicts of interest. }

%%%%%%%%%%%%%%%%%%%%%%%%%%%%%%%%%%%%%%%%%%
%% optional
\abbreviations{Abbreviations}{The following abbreviations are used in this manuscript:\\

\noindent
\begin{tabular}{@{}ll}
NIST & National Institute of Standards and Technology\\
%IC & Imperial College\\
VAMDC & Virtual Atomic and Molecular Data Centre\\
NLTE & non-local thermodynamic equilibrium\\
LLmodels & Line-by-line opacities model atmospheres\\
\end{tabular}}

%=====================================
% References, variant A: internal bibliography
%=====================================
\begin{adjustwidth}{-\extralength}{0cm}
	\reftitle{References}
%	\externalbibliography{yes}
%	\bibliography{paper.bib}

\end{adjustwidth}

%%%%%%%%%%%%%%%%%%%%%%%%%%%%%%%%%%%%%%%%%%
%% optional
%\sampleavailability{Samples of the compounds ...... are available from the authors.}

%%%%%%%%%%%%%%%%%%%%%%%%%%%%%%%%%%%%%%%%%%
\end{document}